\documentclass[10pt,twocolumn]{article}
\usepackage[margin=0.9in,columnsep=0.25in]{geometry}
\usepackage[T1]{fontenc}
\usepackage{mathptmx}
\usepackage{graphicx}
\usepackage{booktabs}
\usepackage{xcolor}
\usepackage{microtype}
\usepackage[hidelinks]{hyperref}
\usepackage{caption}
\usepackage{titlesec}
\titleformat*{\section}{\large\bfseries}
\titleformat*{\subsection}{\normalsize\bfseries}
\usepackage{enumitem}
\setlist{nosep,leftmargin=1.2em}

\newcommand{\meas}{\textsc{measured}}
\newcommand{\der}{\textsc{derived}}
\newcommand{\proj}{\textsc{projected}}
\newcommand{\env}{\textsc{envisioned}}

\title{\LARGE\bfseries The GPU Needs an Operating System\\[2pt]
\large\normalfont Putting idle AI hardware to work}

\author{%
Reza Nourai \qquad Behnam Bastani\\
\normalsize OpenInfer\thanks{Contact: \texttt{research@openinfer.io}.}\\
\normalsize \texttt{openinfer.io}}

\date{}

\begin{document}
\maketitle

\begin{abstract}
AI chips sit mostly idle while serving models, because replies must be fast and each step waits on memory. On two large open models, the math units were busy only about 6--11\% of the time. We argue the GPU needs an operating system that keeps urgent work on schedule and fills idle capacity with other useful work. A feature that ships today recovers a quarter of a GPU with work designed to share, and a tenth with real jobs, at 5\% slower decode. Our research system does better: one resident kernel runs a whole model faster than vLLM's best for a single request, and shares the GPU at under 2\% slowdown. Scaling beyond one hand-built model needs compilers and a few new chip features, which we specify.
\end{abstract}

\section{Introduction}

The chips that run AI are expensive, scarce and power-hungry, yet most of the time they are barely working. To reply quickly, a GPU works on only a few requests at a time. Each request bottlenecks on memory, leaving the math units idle. On a large, widely used open model, the math units were busy only about 7\% of the time. We pay for the whole chip and use a fraction of it.

Computers solved this decades ago with the operating system. A laptop runs a browser, music and a backup at once because the operating system shares the processor and memory and lets urgent work go first. GPUs have nothing like it: each AI program assumes it owns the whole chip. We argue the GPU needs its own operating system, one that keeps urgent work, like a person waiting for an answer, on schedule and fills idle time with useful work such as batch jobs, training or evaluation.

\paragraph*{Why we wrote this.} OpenInfer is an R\&D lab working on the systems problems that keep AI from running as efficiently as its hardware allows. Idle accelerators are one of the first and largest of those problems.

The paper makes the case in seven steps.
\begin{enumerate}
\item \textbf{Today's systems waste the hardware} (\S\ref{sec:idle}). How idle GPUs are when serving large models at realistic response times, and why.
\item \textbf{The future is an operating system for the GPU} (\S\ref{sec:os}). What it would manage, and how it would decide what runs when.
\item \textbf{It cannot simply be bolted on} (\S\ref{sec:kernels}). Today's AI programs are not written to share a chip. Work must be split into pieces that can pause and resume, and compilers, not people, should do the splitting.
\item \textbf{What today's features achieve} (\S\ref{sec:eval}). A feature that ships today puts a good part of the idle capacity to work, but real off-the-shelf jobs recover much less, which shows where the effort must go.
\item \textbf{What we have built} (\S\ref{sec:built}). A research system that runs a whole model in one resident kernel and shares the GPU at a fraction of the cost.
\item \textbf{What needs new chips} (\S\ref{sec:limits}). Some limits software cannot fix. We ask chip makers for three features: fast pausing of background work, priority for urgent work in memory, and partitions the hardware enforces.
\item \textbf{Where this goes} (\S\ref{sec:outlook}). When chips, software and models are designed together for sharing, far more of what we pay for gets used.
\end{enumerate}

The body is written for technical readers who are not GPU specialists; the appendix gives every number and its source.

\section{Today's systems do not use the hardware}
\label{sec:idle}

A GPU serving a large model at interactive speed mostly waits for memory, and the cause is structural.

A GPU is built from many small processors called \emph{SMs} (188 on the NVIDIA RTX PRO 6000 Blackwell we used for the measurements in this paper), each with \emph{tensor cores} for matrix math. Generating text one token at a time is \emph{decode}; for every token it reads the model's weights and the \emph{KV cache}, the stored context of the conversation, from memory. At small batches there is little math per byte read, so the tensor cores finish quickly and wait.

Mixture-of-experts (MoE) models, where each token uses only a few of many experts, make this worse. We served a 397B-parameter MoE with stock vLLM on four RTX PRO 6000 GPUs under an agent workload, at the highest load that met 97\% of its deadlines; its tensor cores were busy about 7\% of the time, and 2.9\% in a pool that only decodes. Kimi K2, with about a trillion parameters, sends each token to 8 of 384 experts, so a batch of 256 gives each expert about five tokens. On these GPUs K2 met its deadlines under neither of two standard deployments: eight 96\,GB cards hold its 554\,GB of weights and leave little room for the KV cache. At the loads it did carry, its tensor cores were busy 6--11\% of the time (Appendix~\ref{app:a1}).

\begin{figure}[t]
\centering
\includegraphics[width=\columnwidth]{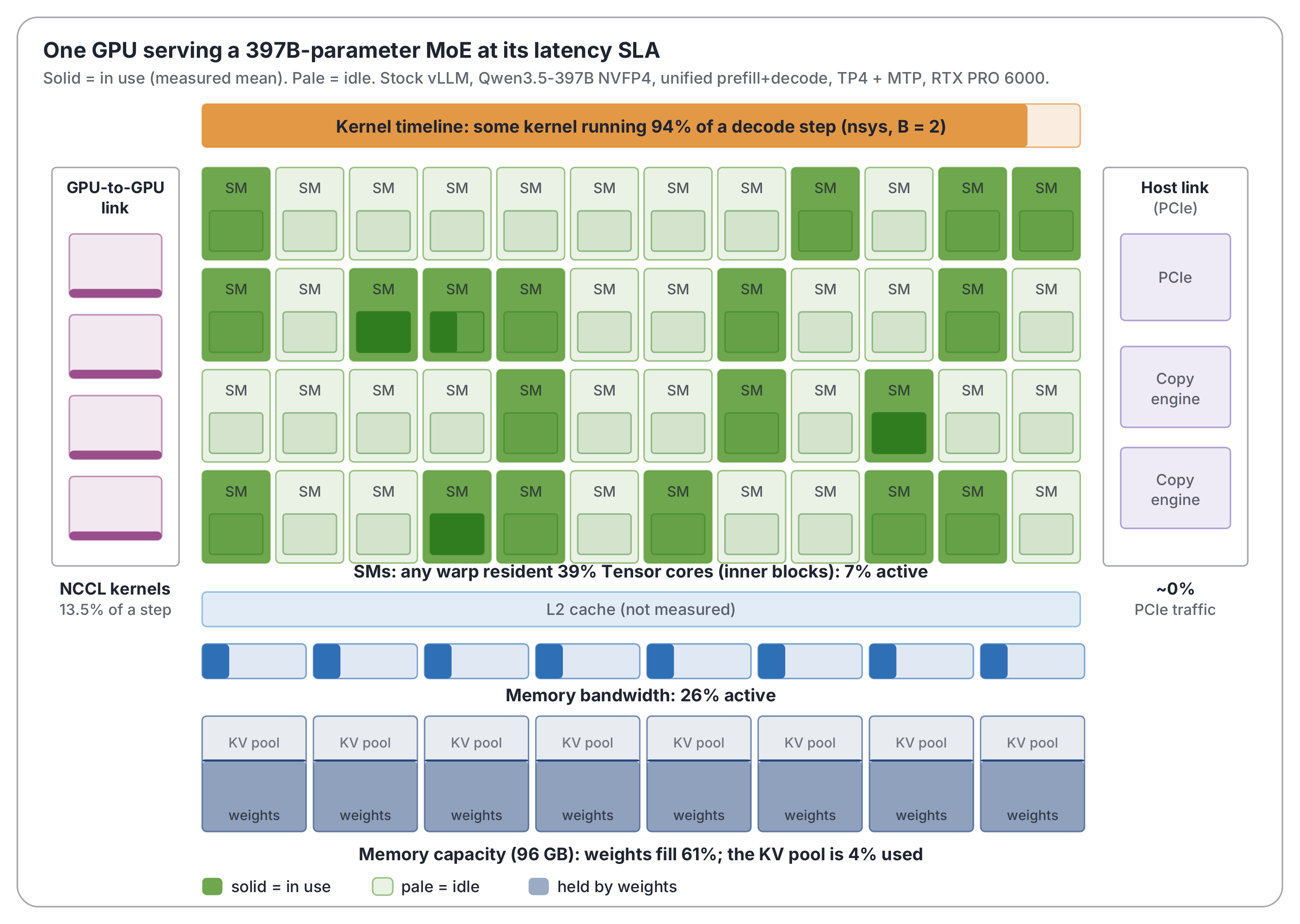}
\caption{One GPU serving the 397B model at the highest load that met its deadlines; each resource is solid where in use and pale where idle. Conceptual layout, not to scale: 48 SM tiles stand for the chip's SMs, and shaded counts match the measured fractions (placement is illustrative). Means over the measured window; tensor p90 is 19\%. Copy engines have no counter on this box, so the host link is shown as PCIe traffic; GPU-to-GPU traffic goes over PCIe here (NVLink on HGX systems). Appendix~\ref{app:a1}.}
\label{fig:chip}
\end{figure}

\textbf{Why this is structural.} A chip's \emph{ridge} is the ratio of math it can do to bytes it can read from memory. Chips fed from HBM or GDDR, including NVIDIA and AMD GPUs and Google TPUs, have ridges of roughly 240 to 560 operations per byte. Decode at batch $B$ supplies only about $B$ operations per byte of weights (2$B$ for 8-bit weights), and for an MoE, $B$ is the handful of tokens each expert sees. The math units are then at most a few times $B \div \mathrm{ridge}$ busy, around 1\% at batch 2, whether they are tensor cores or a TPU's systolic array (Appendix~\ref{app:a1}). Chips built around large on-die SRAM read tens of times faster, so their ridge falls to single digits and they run short of capacity instead.

\section{The future is an OS-like system}
\label{sec:os}

\begin{figure}[t]
\centering
\includegraphics[width=\columnwidth]{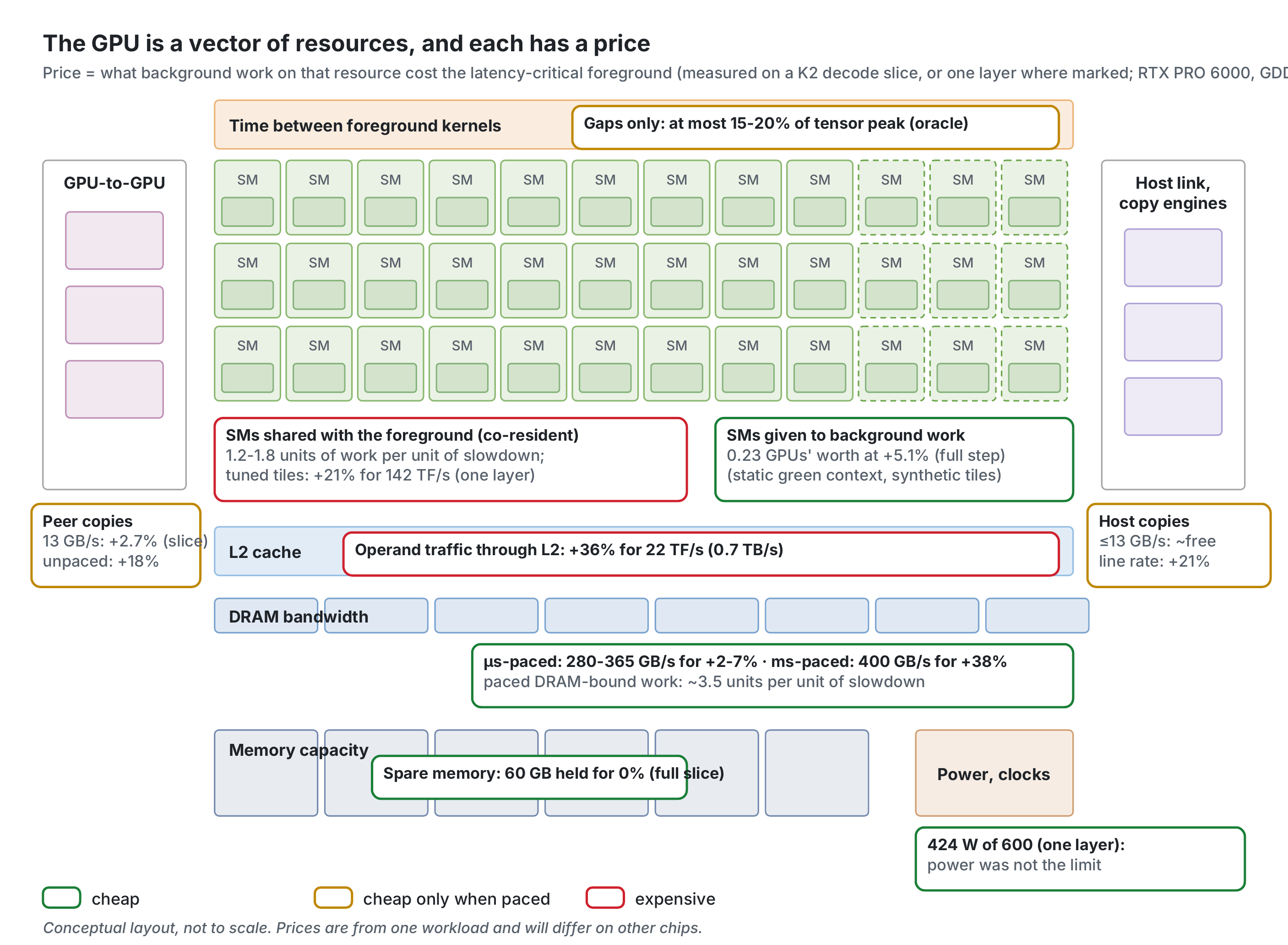}
\caption{The same chip, each resource tagged with its measured price (Appendix~\ref{app:a2}).}
\label{fig:resources}
\end{figure}

An operating system lets one computer run urgent and background work together. A GPU has nothing like it: it runs queued work in order, and nothing running can be asked to leave. An OS would treat the chip as separate resources, each with its own cost: SM time, tensor cores, memory bandwidth, the on-chip cache (L2), memory capacity, data links and power (Figure~\ref{fig:resources}). They trade at very different rates: holding 60\,GB of idle memory for another job cost the urgent work nothing measurable, while extra compute beside it cost a lot (Appendix~\ref{app:a2}).

The OS spends \emph{slack}, the margin by which urgent work beats its deadline, on background work. Our budget is 6\% on the slow tail: a step is on time if it takes no more than 1.06 times the urgent work's 99th-percentile step time when it runs alone.

Three of today's mechanisms fall short:
\begin{itemize}
\item \textbf{Stream priorities do not interrupt running work.} A low-priority job in 1\,ms pieces slowed urgent work up to 10$\times$. On an AMD MI300X, urgent work waited 180\,ms behind a 200\,ms background job.
\item \textbf{The length of each background piece decides the cost.} Pieces of 20\,\textmu s or less were tolerable; at 50\,\textmu s the urgent work missed every deadline (Appendix~\ref{app:a2}).
\item \textbf{Partitions are coarse or fixed.} A \emph{green context} is an NVIDIA feature that reserves a set of SMs for one queue of work. It is the best tool we found, but fixed per queue. MIG splits the chip into up to four isolated instances, and re-splitting needs a reset.
\end{itemize}

The proposed OS has six layers, deciding every few milliseconds at the top and enforcing every few microseconds on the GPU at the bottom:
\begin{description}[style=unboxed,leftmargin=0pt]
\item[L0, chip features:] signals, queues, fast pausing, memory priority (fragments exist today).
\item[L1, kernel rules:] short, restartable pieces of work that state what they use.
\item[L2, on-GPU scheduler:] places those pieces on SMs.
\item[L3, resource governor:] paces memory, cache and copy traffic; lends spare memory.
\item[L4, cost model:] predicts how much background work will slow urgent work.
\item[L5, admission:] chooses background work within the slack.
\end{description}
We have built parts of L1--L3; L4 and L5 are \env{}.

\section{What has to change in kernels and compilers}
\label{sec:kernels}

A chip can only be shared if its programs, called kernels, are written to be shared. Today's kernels assume they own every SM and will run to completion. Kernels for a shared GPU would run in pieces of tens of microseconds, stop cleanly between pieces when asked, and keep their progress in memory. They would state what they use, especially bytes read per unit of math, so the OS can price them, and background work would reuse data already in the cache.

One requirement is now measured: urgent work can only give SMs away if it stays fast on fewer of them. A well-written streaming kernel reaches 99\% of the card's memory bandwidth on 24 of its 188 SMs, provided each SM keeps about 100\,KB of reads in flight (Appendix~\ref{app:a3}). Today's decode kernels reach about 50\% of bandwidth on the whole chip and slow by 22\% when one phase is squeezed onto 60 SMs.

\begin{figure}[t]
\centering
\includegraphics[width=\columnwidth]{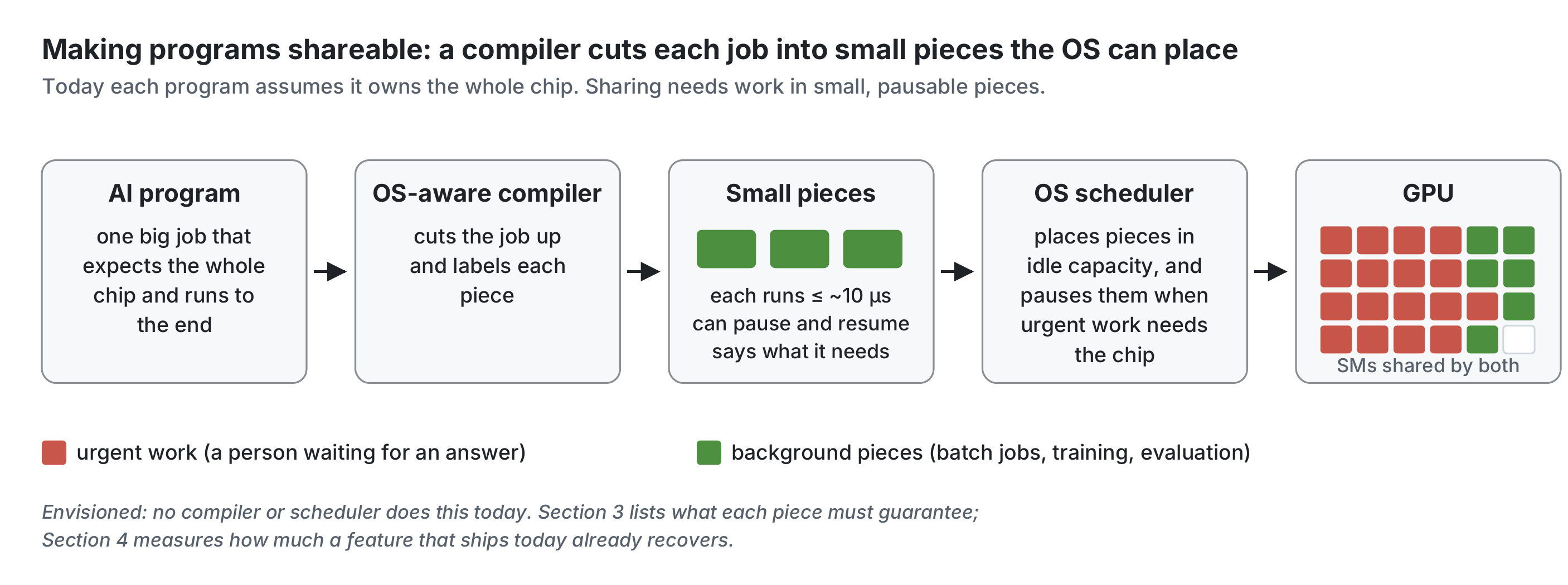}
\caption{How programs would become shareable. A compiler cuts each job into small pieces that can pause, and an OS scheduler places them in idle capacity. This is \env{}; Appendix~\ref{app:a4} describes the engineering pieces we have built separately.}
\label{fig:osmode}
\end{figure}

Nobody will hand-write thousands of such kernels, so compilers such as Triton and libraries such as CUTLASS should generate them: from one source, versions sized to the share of the chip the OS offers, each with a time limit, safe stopping points and a list of the resources it uses. They are not there yet: asking NVIDIA's matrix-multiply library to target a partition made our step slower (Appendix~\ref{app:a3}).

Model architecture can help too. An MoE with evenly sized experts and balanced routing, or a hybrid that replaces most full-attention layers with linear attention such as Kimi Delta Attention, runs fewer, more uniform kernels whose cost does not grow with context, giving a scheduler longer, more predictable gaps. Models that produce their own cache-friendly background work, such as adapter training or reward scoring, share better than unrelated jobs. These directions are untested.

\section{What co-running achieves today}
\label{sec:eval}

We measured how much extra work a GPU can do beside a real model's decode step (co-running) with features that ship today; most of the limit is in today's kernels. The urgent work was a proxy for one GPU's share of serving Kimi K2: all 61 layers, batch 8 at 32k tokens of context, on one RTX PRO 6000. A static green context reserved part of the chip for it, and background work ran on the rest (Appendix~\ref{app:a4}).

With synthetic background work designed to share, which keeps its data in the cache, tensor-core activity rose from 8.7\% to about 30\% while median steps got 5.1\% slower, inside the 6\% budget. That is about a quarter of a GPU's worth of extra work, the ceiling of this mechanism; across a 64-GPU deployment, waiting for the slowest GPU would push the tail just over budget. Real off-the-shelf work recovers less: a tuned matrix-multiply job (NVIDIA's library, 8-bit precision) harvested about 0.09 of a GPU inside the budget, with steps 4.7\% slower and tensor activity at 15\%. Pushing it to 0.13 slowed steps 9.9\%, well over budget (Appendix~\ref{app:a4}).

\begin{table}[t]
\centering\footnotesize
\caption{Full-step co-run on the K2 proxy (\meas{}; six samples per configuration; Appendix~\ref{app:a4}). ``Worth'' is background rate over that work's best solo rate on this GPU.}
\label{tab:fullstep}
\begin{tabular}{p{3.0cm}rrrr}
\toprule
configuration & p50 & p99 & attain & worth \\
\midrule
tiles, FG on 172 SMs & +0.3\% & +0.5\% & 100\% & 0.06 \\
tiles, FG on 156 SMs & +4.7\% & +4.7\% & 100\% & 0.15 \\
tiles, FG on 140 SMs & +5.1\% & +5.5\% & 95--100\% & 0.23 \\
FP8 GEMM $1536^3$, FG 168 & +4.7\% & +4.8\% & 99.8\% & 0.09 \\
FP8 GEMM $2048^3$, FG 156 & +9.9\% & +9.8\% & 0--32\% & 0.13 \\
per-phase partitions + doorbell & +11.8\% & +16.8\% & 3--22\% & 0.41 \\
\bottomrule
\end{tabular}
\end{table}

\begin{figure}[t]
\centering
\includegraphics[width=\columnwidth]{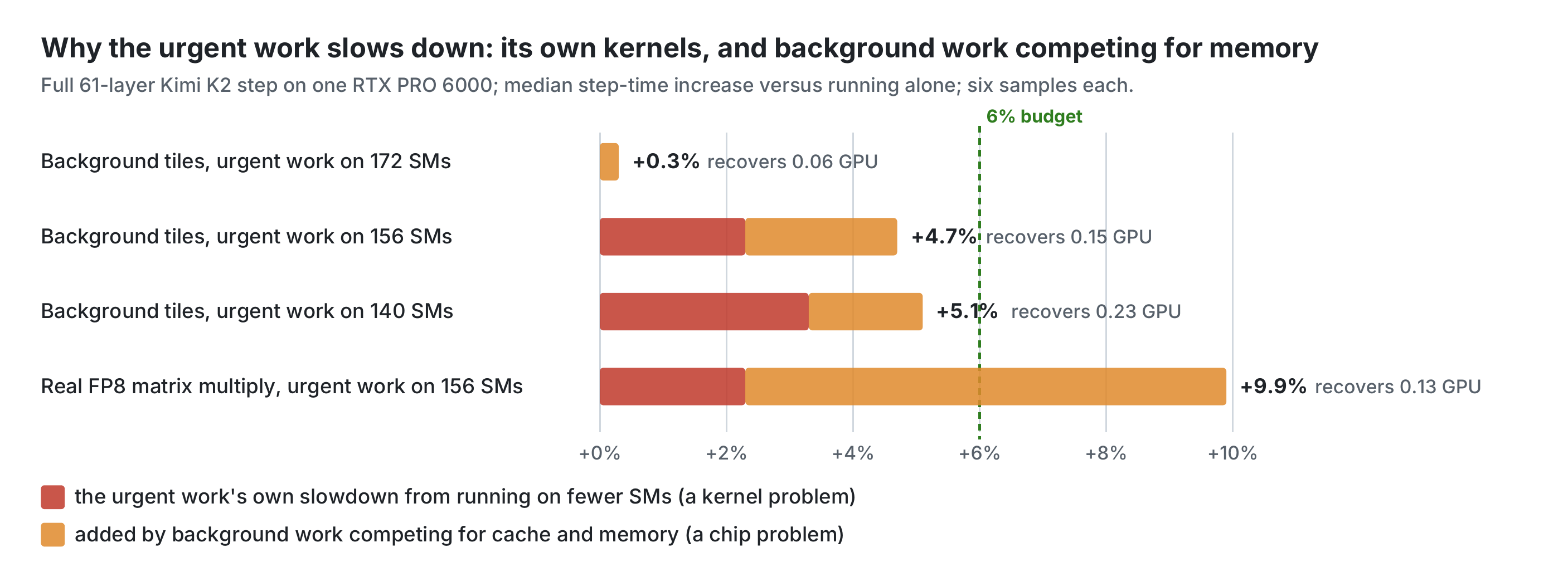}
\caption{Each co-run's slowdown split into its two causes: the urgent work's own loss from running on fewer SMs, and the extra added by background work competing for cache and memory (Appendix~\ref{app:a4}).}
\label{fig:cost}
\end{figure}

The urgent work's own kernels cause most of the cost: each SM it gives away is lost almost one for one. Alone it uses about half the memory bandwidth, so it could run about twice as fast in principle. Real background jobs add a second cost, competing for the cache and memory in a way no shipped feature controls at runtime. A scheme that changed the partition at every phase of the step looked better in an early one-layer test and worse on the full step (Appendix~\ref{app:a4}).

\section{What we have built}
\label{sec:built}

We have built two pieces of this system as working research systems on the same GPU.

\textbf{A model in one resident kernel.} A single persistent kernel runs all of Llama-3.1-8B for the life of the process. It admits new requests itself, samples and checks stop conditions on the GPU, and streams tokens to the CPU through a memory ring, so the CPU is never in the decode loop. For a single request it already decodes 3\% faster than vLLM's best configuration on the same weights (Appendix~\ref{app:a7}).

\textbf{Background work inside that kernel.} Giving some of that kernel's SMs to background work that pauses at phase boundaries recovered as much of the GPU as the best shipped partition, at a fraction of the cost:

\begin{table}[h]
\centering\footnotesize
\begin{tabular}{p{3.9cm}rr}
\toprule
 & background & urgent work's \\
 & recovered & slowdown \\
\midrule
shipped partition (\S\ref{sec:eval}) & 0.23 of a GPU & +5.1\% \\
resident kernel, 1 request & 0.22 of a GPU & +0.3\% \\
resident kernel, 16 requests & 0.24 of a GPU & +1.6\% \\
\bottomrule
\end{tabular}
\end{table}

\section{What limits co-running today}
\label{sec:limits}

Co-running is limited first by today's urgent-work kernels and then by a few missing chip features.

\begin{figure}[t]
\centering
\includegraphics[width=\columnwidth]{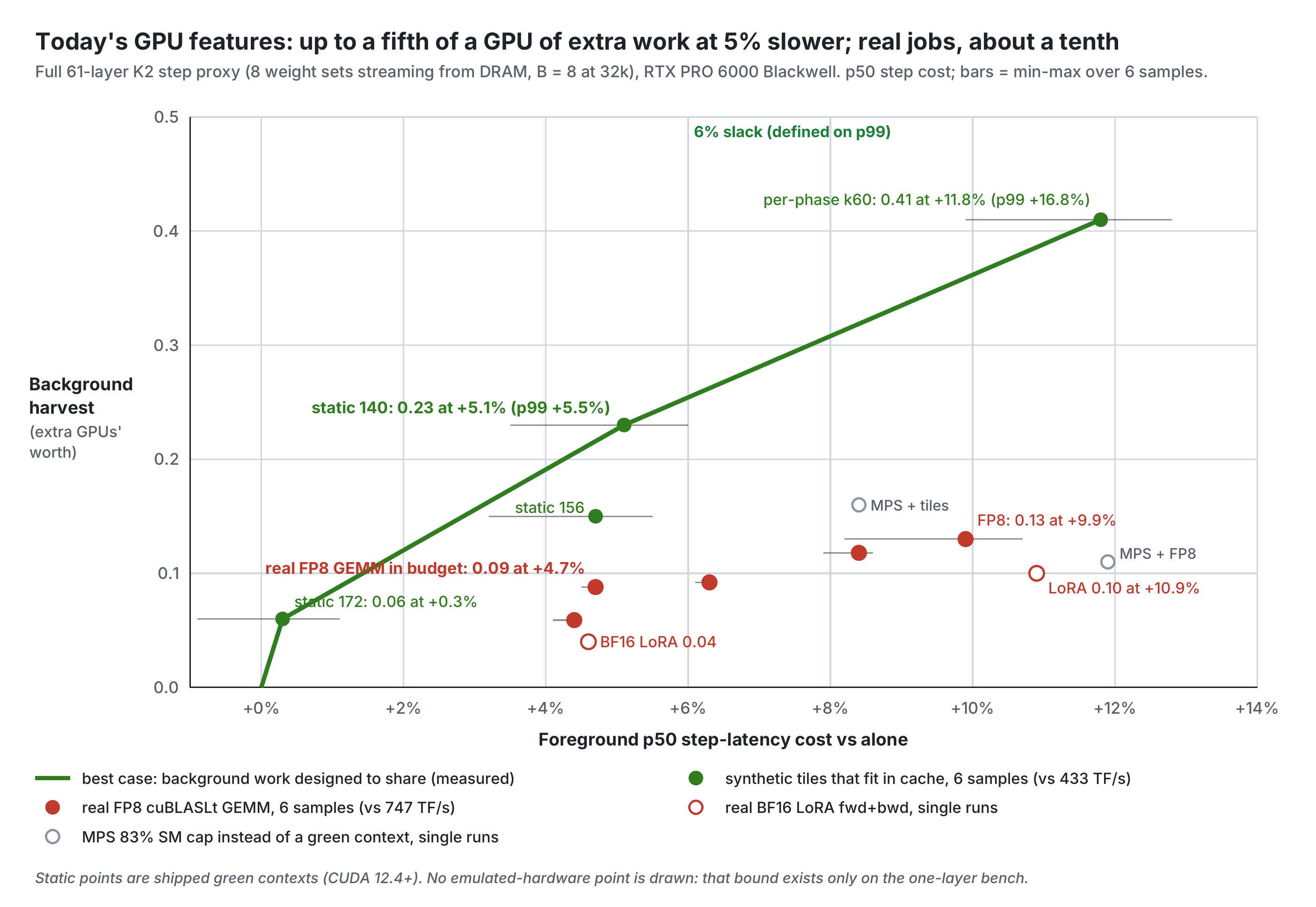}
\caption{Extra work beside K2 decode, as a fraction of a whole GPU's output (vertical), against how much slower the median decode step gets (horizontal). Each point is a measured green-context configuration on the full 61-layer step; the green line connects the best, and the dashed line marks the 6\% budget, defined on the slowest 1\% of steps. The gap between the line and the real jobs is the cache and memory interference the requests below address (Appendix~\ref{app:a4}).}
\label{fig:tradeoff}
\end{figure}

Kernels that reached full bandwidth on a partition would free more of the chip at the same cost and leave more slack (\S\ref{sec:kernels}). Interference in the cache and memory needs the chip's help.

On an AMD MI300X, CU masks (AMD's equivalent of green contexts) split the chip cleanly, but priorities did not interrupt running work, and stopping background work by ending and restarting it was costly (Appendix~\ref{app:a5}).

\textbf{Requests to chip makers.} Software alone cannot remove these costs; we would welcome working with GPU and accelerator makers on them. Each request is tied to a measured cost (Appendix~\ref{app:a6}):
\begin{description}[style=unboxed,leftmargin=0pt]
\item[S1. Fast pausing.] Pause background work within a guaranteed 5--20\,\textmu s and resume it without starting over. Neither vendor lets ordinary programs do this today.
\item[S2. Priority in the cache and memory, set while running.] Tag work by kind and let urgent work go first in the cache and memory controllers. MIG isolates memory only between fixed instances.
\item[S3. Per-kernel partitions that the hardware honours.] CUDA accepts a partition for each kernel in a recorded sequence but ignores it; working around that cost us 1.4\,ms per step.
\item[S4. Priority inside an SM,] so urgent work's instructions run first beside background work.
\item[S5--S7. Supporting features:] fast chip-wide signals and queues, detailed counters that ordinary programs can read, and priority and rate limits for data copies and power.
\end{description}

\section{Outlook}
\label{sec:outlook}

The biggest gains come when models, compilers, runtime and chips are designed together for sharing.

\begin{figure}[t]
\centering
\includegraphics[width=\columnwidth]{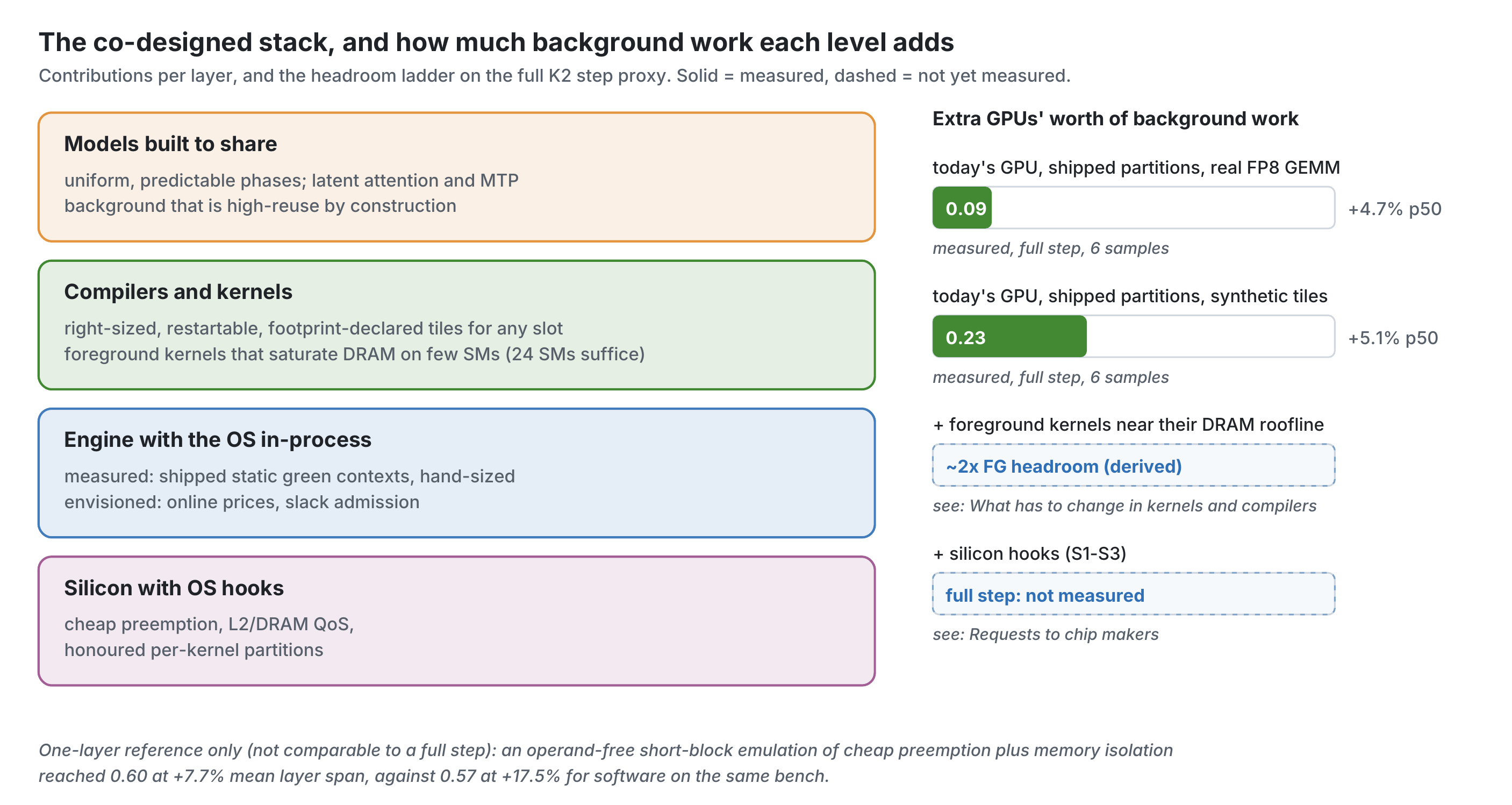}
\caption{The co-designed stack and how much background work each level adds, measured where we have data and left open where we do not (Appendix~\ref{app:a6}).}
\label{fig:stack}
\end{figure}

Models would produce even work and cache-friendly background tasks; compilers would generate kernels that pause cleanly and stay fast on part of the chip; a runtime would size partitions and admit work within the slack; chips would add the requested features. We expect the results on pausing and partitioning to carry over to datacenter GPUs, and the size of the memory interference and the 24-SM point at which bandwidth saturates to differ. The same approach should carry over to other accelerators, and we plan to test it there.

\section{Related work}
\label{sec:related}

\textbf{Slack-driven co-location on CPUs.} Heracles~\cite{heracles} and PARTIES~\cite{parties} treat slack as the budget and the server as a vector of resources.

\textbf{Background work beside live serving.} ROSE~\cite{rose}, LeMix~\cite{lemix}, ConServe~\cite{conserve}, HyGen~\cite{hygen}, MOLT~\cite{molt} and LayerLoom~\cite{layerloom}. LayerLoom, which runs LoRA backward work within the slack of in-flight requests, is the closest prior instance of our slack policy, for one pair of workloads. CrossPool~\cite{crosspool} serves rarely used MoE models from shared weight and KV-cache pools, with persistent kernels polling per-model queues. RL frameworks (veRL/HybridFlow, RLHFuse, StreamRL, AReaL, PipelineRL)~\cite{rlframeworks} overlap their own phases, with no response-time target.

\textbf{Spatial sharing and right-sizing.} KRISP~\cite{krisp} (kernel-wise right-sizing), ECLIP~\cite{eclip} (energy-efficient co-location on partitioned GPUs), LithOS~\cite{lithos}, Bullet~\cite{bullet}, SGLang's PD-mux (green contexts shared by prefill and decode), MuxWise~\cite{muxwise} (intra-GPU prefill--decode multiplexing), Vitamin-E~\cite{vitamine} (formerly DetShare: spatial sharing on green contexts with central dispatch, lower inference tail latency beside training) and MuxServe~\cite{muxserve} (spatial-temporal multi-model serving over MPS). Our headline mechanism is close to these.

\textbf{Preemption and yield.} Tanasic et al.~\cite{tanasic}, Chimera~\cite{chimera}, EffiSha~\cite{effisha}, FLEP~\cite{flep}, REEF~\cite{reef}, Tally~\cite{tally}, GPreempt~\cite{gpreempt}, XSched~\cite{xsched} and ExpertPlex~\cite{expertplex}. Hummingbird~\cite{hummingbird} builds microsecond-scale, deadline-aware preemption on closed-source GPUs. Our block-lifetime measurements show how short background work must be beside MoE decode; S1 asks for that bound in hardware.

\textbf{Intra-SM sharing and fusion.} SMK~\cite{smk}, Warped-Slicer~\cite{warpedslicer}, Wang et al.~\cite{wangqos}, HFuse~\cite{hfuse}, POD-Attention~\cite{podattn}, NanoFlow~\cite{nanoflow}, elastic kernels~\cite{pai} and megakernels~\cite{mpk,hazy}. GPUOS~\cite{gpuosyang} shares our title only: a persistent kernel that runs NVRTC-injected operations from a host-managed queue to cut launch overhead, without deadline classes, slack or per-resource scheduling. Our OS would use that mechanism.

\textbf{Memory partitioning.} SGDRC and MIG; Orion~\cite{orion} schedules by interference class.

We add measured prices for background work beside large-MoE decode, the block-lifetime limit, the full-step tail cost of shipped partitions with synthetic and real work (including the wait for the slowest of many GPUs), and an attribution of what remains to kernels, memory contention and partition mechanics, on two vendors.

\appendix
\section{Methods and measurements}
\label{app:methods}

Labels: \meas{} (OpenInfer research systems), \der{} (arithmetic on measured or vendor figures), \proj{}, \env{}. Measured numbers come from our research systems on one card and are not upper limits. Sources are our measurement logs. All GPU measurements are on NVIDIA RTX PRO 6000 Blackwell (sm\_120: 188 SMs, 96\,GB GDDR7, 128\,MB L2, no NVLink, no tcgen05 or TMEM) unless marked MI300X.

\subsection{Serving counters and the ridge table}
\label{app:a1}

\textbf{Qwen3.5-397B} (\S\ref{sec:idle}, Figure~\ref{fig:chip}). NVFP4, stock vLLM v0.32, unified prefill+decode, TP4 with two-token MTP, four GPUs, 30-minute agent window, CUPTI PM counters at 10\,kHz. Held point: 97.2\% of calls on time at 0.02 calls/s; at 0.025 attainment fell to 91.5\%.

\begin{table}[h]
\centering\footnotesize
\begin{tabular}{p{3.3cm}p{3.6cm}}
\toprule
resource (mean active, held point) & Qwen3.5-397B \\
\midrule
tensor pipes & $\sim$7\% (p90 19\%); 2.9\% in a decode-only pool \\
SMs (any warp resident) & $\sim$39\% \\
DRAM bandwidth & $\sim$26\% \\
PCIe & near zero \\
KV cache capacity used & $\sim$4\% \\
weights, share of 96\,GB & $\sim$61\% (\der{}: 234\,GB over 4 GPUs) \\
kernel busy, decode step (nsys, $B{=}2$) & $\sim$94\% (compute 80.4\%, NCCL 13.5\%) \\
NCCL share of a decode step & 13.5\% (all-reduce 11.9\% + all-gather 1.6\%) \\
\bottomrule
\end{tabular}
\end{table}

\textbf{Kimi K2} (\S\ref{sec:idle}). Kimi-K2-Thinking, INT4 as published (554\,GB); stock vLLM v0.32 on eight GPUs; same window and counters. \emph{No held point under either deployment.} The limit is memory: 554\,GB of weights on $8\times96$\,GB, with BF16 KV because FP8 KV fails on sm\_120 (vLLM's Triton MLA decode, its only MLA backend there, needs 102{,}400\,B of shared memory; the limit is 101{,}376\,B). Misses are KV-capacity evictions that force long re-prefills. Deadlines: each deployment's idle latency $\times 2 \times 1.25$, as for the 397B.

\begin{table}[h]
\centering\footnotesize
\begin{tabular}{p{2.5cm}p{1.3cm}rrr}
\toprule
deployment & KV pool & rate & attain & tensor \\
\midrule
TP8 over PCIe (latent KV replicated on every rank) & $\sim$182k tok & 0.0025 & 66\% & 6.1\% (23\%) \\
TP8 & & 0.005 & 54\% & 6.8\% (25.6\%) \\
TP4 $\times$ DP2 + EP8 (KV sharded) & 149.5k per rank, 1.64$\times$ TP8 & 0.005 & 16.7\% & 10.7\% (19.4\%) \\
\bottomrule
\end{tabular}
\end{table}

Each DP rank in the third row holds the 21.8\,GiB of non-expert weights. TP1 $\times$ DP8 does not fit ($\sim$88\,GiB of weights per GPU, \proj{}); TP2 $\times$ DP4 leaves no KV room at chunk 8192. TP4 $\times$ DP2 needs chunk 2048 and runs EP collectives over PCIe (idle TPOT 27.9\,ms against 19.1 for TP8).

\textbf{Ridge table} (\der{} from vendor specifications; 16-bit dense peaks). Decode at batch $B$ does about $B$ FLOP per byte of 16-bit weights (2$B$ for 8-bit), ignoring KV traffic; for an MoE, $B$ is the per-expert batch.

\begin{table}[h]
\centering\footnotesize
\begin{tabular}{p{1.9cm}rrrr}
\toprule
machine & peak & bandwidth & ridge & $B{=}2$ \\
\midrule
RTX PRO 6000 Blackwell & $\sim$0.5\,PF\textsuperscript{a} & $\sim$1.79\,TB/s GDDR7 & $\sim$280 & $\sim$1\% \\
AMD MI300X & $\sim$1.3\,PF & 5.3\,TB/s HBM & $\sim$250 & $\sim$1\% \\
Google TPU v5e & 197\,TF & $\sim$0.82\,TB/s & $\sim$240 & $\sim$1\% \\
Google TPU v6e & 918\,TF & $\sim$1.6\,TB/s & $\sim$560 & $<$1\% \\
SRAM dataflow (Groq LPU class) & $\sim$190\,TF\textsuperscript{b} & $\sim$80\,TB/s SRAM\textsuperscript{b} & $\sim$2--3 & balance \\
Wafer-scale (Cerebras WSE-3) & $\sim$125\,PF\textsuperscript{b,c} & $\sim$21\,PB/s SRAM\textsuperscript{b} & $\sim$6 & $\sim$1/3 \\
\bottomrule
\end{tabular}

\smallskip\raggedright\footnotesize \textsuperscript{a}third-party sheet. \textsuperscript{b}vendor. \textsuperscript{c}may be sparse.
\end{table}

\subsection{Resource prices, slack and block lifetime}
\label{app:a2}

\textbf{Resources} (\S\ref{sec:os}):

\begin{table}[h]
\centering\footnotesize
\begin{tabular}{p{2.0cm}p{1.9cm}p{2.6cm}}
\toprule
resource & unit & shared today by \\
\midrule
SM issue slots, warp contexts & warp-\textmu s per SM & first come, first resident; static SM caps (MPS, green contexts) \\
tensor pipes & MMA-\textmu s & residency, implicitly \\
DRAM and L2 bandwidth & bytes per \textmu s, queueing delay & nothing dynamic; static MIG instances \\
device memory capacity & bytes $\times$ time, revocation cost & an allocator, no leases \\
copy engines, PCIe, NVLink & bytes per \textmu s per direction & driver order \\
power and clocks & joules, and the clock cost to others & board-wide, invisibly \\
time between kernels & \textmu s gaps & nobody \\
\bottomrule
\end{tabular}
\end{table}

\textbf{Prices in Figure~\ref{fig:resources}} (\meas{}; foreground is a full K2 decode slice at a 50\,ms SLO unless marked):

\begin{table}[h]
\centering\footnotesize
\begin{tabular}{p{3.6cm}p{3.3cm}}
\toprule
resource used by background & foreground cost \\
\midrule
spare memory held & 60\,GB at 0\% \\
host copies, paced / unpaced & $\leq$13\,GB/s at $\sim$0\%; 90\,GB/s at +21\% \\
peer copies, paced / unpaced & 13\,GB/s at +2.7\%; 51\,GB/s at +18\% \\
DRAM traffic, \textmu s-paced (in-kernel token bucket) & 280--365\,GB/s at +2--7\% (three seeds) \\
DRAM traffic, ms-paced (host bursts) & 400\,GB/s at +38\% \\
operand traffic through L2 (real-operand GEMM tile, 5\,\textmu s blocks) & +36\% for 22\,TF/s (0.69\,TB/s) \\
co-resident SM work & 1.2--1.8 units of work per unit of slowdown \\
time between foreground kernels only & at most $\sim$15--20\% of tensor peak (oracle) \\
SMs given to background (full step, green context) & 0.23 GPUs' worth at +5.1\% (\ref{app:a4}) \\
power & 424\,W of 600 under co-run (one layer); not binding \\
\bottomrule
\end{tabular}
\end{table}

Paced DRAM-bound work bought $\sim$3.5 units per unit of slowdown at 100\,GB/s and 1.8 at 400\,GB/s. Loads add super-linearly: DRAM 100\,GB/s (+4\%), peer 13\,GB/s (+2.7\%) and host 13\,GB/s ($\sim$0\%) together cost +14\%.

\textbf{Slack.} A step is on time if it takes $\leq 1.06 \times$ the foreground-alone p99 (the 6\% budget); attainment is the share of steps on time.

\textbf{Priorities} (\S\ref{sec:os}). Low-priority background in 1\,ms blocks slowed the foreground up to 10$\times$ (\meas{}, design-phase spike). MI300X: Appendix~\ref{app:a5}.

\textbf{Block lifetime} (\S\ref{sec:os}). Background is operand-free register MMA in blocks of $N$\,\textmu s, an upper bound; full K2 slice; ``met'' is attainment of the 50\,ms SLO, a looser test than the 6\% budget.

\begin{table}[h]
\centering\footnotesize
\begin{tabular}{p{1.2cm}p{3.1cm}rr}
\toprule
$N$ & foreground step & met & BG MMA \\
\midrule
2--20\,\textmu s & +5.6\% to +9.5\% (5\,\textmu s, three seeds: +6, +9.5, +9\%) & 92--100\% & 158--264\,TF/s \\
50\,\textmu s & +47\% & 0\% & 316\,TF/s \\
200\,\textmu s (spin) & +165\% & 0\% & \\
\bottomrule
\end{tabular}
\end{table}

The knee is 20--50\,\textmu s for operand-free blocks; with real operands, 5\,\textmu s blocks already cost +36\%. Yielding by relaunch costs the background 41--44\% of its throughput at 2--5\,\textmu s.

\textbf{Built parts of L1--L3} (\S\ref{sec:os}): persistent background tiles with a yield flag, a device-memory doorbell, and \textmu s DRAM pacing.

\subsection{Bandwidth vs SM count and the foreground roofline}
\label{app:a3}

\textbf{DRAM bandwidth vs SM count} (\S\ref{sec:kernels}). 4\,GiB read-only buffer, 3 runs per point, each $k$-SM run on a $k$-SM green context. ``ld'' is a vectorised load stream with $\sim$96\,KB in flight per SM; ``TMA'' is a bulk-copy streamer.

\begin{table}[h]
\centering\footnotesize
\begin{tabular}{rrrrr}
\toprule
SMs & ld GB/s & \% peak & TMA GB/s & \% peak \\
\midrule
8 & 569 & 37\% & 488 & 32\% \\
16 & 1{,}113 & 72\% & 946 & 61\% \\
24 & 1{,}527 & 99.2\% & 1{,}366 & 89\% \\
32 & 1{,}536 & 99.7\% & 1{,}532 & 99.5\% \\
64 & 1{,}538 & 99.9\% & 1{,}538 & 99.9\% \\
188 & 1{,}538 & 100\% & 1{,}537 & 99.8\% \\
\bottomrule
\end{tabular}
\end{table}

The measured peak, 1{,}538\,GB/s, is 86\% of the 1.79\,TB/s specification. Little's law: $\sim$1.5\,TB/s at the observed latency needs $\sim$2.3\,MB outstanding, about 24 SMs at $\sim$96\,KB each.

\textbf{Foreground roofline} (\S\ref{sec:kernels}--\ref{sec:eval}; full K2 step proxy, alone). 28.9\,GB per step over 37.6\,ms of phase time is 769\,GB/s, 50\% of the 1{,}538\,GB/s peak (an earlier, less tuned measurement read 1.46\,TB/s); the DRAM-roofline step is 18.8\,ms, so $\sim$2$\times$ headroom (\der{}). By phase: experts 14.4\,ms at 61\% of peak, MLA 9.7\,ms at 60\%, dense projections 8.5\,ms at 27\%, shared expert 3.6\,ms at 48\%, router 1.4\,ms at 8\%. At 8--16 tokens cuBLASLt is latency-bound. On 60 SMs the experts phase takes 17.6\,ms (+22\%). Setting \texttt{CUBLASLT\_MATMUL\_DESC\_SM\_COUNT\_TARGET} to the partition size made the step 1--2 points worse.

\subsection{The full-step co-run}
\label{app:a4}

\textbf{Setup} (\S\ref{sec:eval}, Figures~\ref{fig:cost}--\ref{fig:tradeoff}). Foreground: a single-GPU proxy for one EP-64 rank of Kimi K2, 61 layers, eight distinct weight sets streaming from DRAM, batch 8 at 32k context, 200 steps per run, no real all-to-all. Six samples per configuration (three interleaved rounds on two GPUs, after a discarded warm-up), each against the same GPU's alone run (p50 39.0--39.9\,ms). Run-to-run noise is about $\pm$1--1.5 points of p50 cost.

\textbf{GPUs' worth} is background rate divided by the best solo rate of that kind: 433\,TF/s for synthetic tiles (BF16 \texttt{mma.sync}); 747\,TF/s for FP8 cuBLASLt GEMM (the best shape, $8192^3$, a conservative choice; $2048^3$ alone reaches 485); 377\,TF/s for BF16 LoRA. The synthetic tile is a persistent MMA loop over a $\sim$5.5\,MB L2-resident working set with no useful output: an upper bound for high-reuse background work. The EP-64 projection (\proj{}) is the max of 64 draws from each run's per-layer time distribution; correlated ranks would shrink it.

Static green contexts, six samples each, give the p50 range and EP-64 projections for Table~\ref{tab:fullstep}'s rows: $[-0.9,1.1]$ and +0.5/+0.1\%; $[3.2,5.5]$ and +5.3/+4.9\%; $[3.5,6.0]$ and +6.8/+6.6\%; $[8.2,10.7]$ and +10.4/+10.0\%; $[9.9,12.8]$ and +27/+29\%. The 0.06 point (+0.3\%) is inside the noise. At 64 ranks the 140-SM point's p99 (+6.6\%) exceeds the budget; the 156-SM point (+4.9\%) stays inside.

\textbf{Harvest tracks the SMs given away:} 16, 32 and 48 free SMs are 8.5\%, 17\% and 25.5\% of 188, and the harvest is 0.06, 0.15 and 0.23. The foreground alone on 172, 156 and 140 SMs costs $-0.2$, +2.3 and +3.3\% (Figure~\ref{fig:cost}'s ``own loss''; the rest of each bar is the difference of measured means). Figure~\ref{fig:tradeoff} draws the concave hull of all measured points; points between vertices are time-mixes of two configurations, and attainment holds only at the vertices.

\textbf{Real FP8 GEMM inside the budget.} Tuned cuBLASLt FP8 GEMMs back to back on the complement, the best of 11 scouted configurations, six samples each. Green contexts round partitions up (a 172-SM request gives 176/12; 164 gives 168/20).

\begin{table}[h]
\centering\footnotesize
\begin{tabular}{p{2.1cm}rrrr}
\toprule
FG/BG SMs, GEMM & p50 & attain & worth & tensor \\
\midrule
alone (39.01\,ms p50) & & 99.9\% & 0 & 8.7\% \\
176/12, $2048^3\times8$ & +4.4\% & 100\% & 0.059 & 13.1\% \\
168/20, $1536^3\times12$ & +4.7\% & 99.8\% & 0.088 & 15.4\% \\
168/20, $2048^3\times8$ & +6.3\% & 94.7\% & 0.092 & 15.6\% \\
152/36, $1024^3\times32$ & +8.4\% & 26.8\% & 0.118 & 17.5\% \\
\bottomrule
\end{tabular}
\end{table}

In-budget real work is 0.06--0.09 of a GPU (best $\sim$66\,TF/s). $1536^3$ fits better than $2048^3$ on 20 SMs because its working set ($\sim$9\,MB against $\sim$16\,MB) disturbs the foreground's L2 less.

\textbf{Hardware counters, before and after} (\S\ref{sec:eval}; separate runs with device-wide CUPTI counters, steady window, six samples each):

\begin{table}[h]
\centering\footnotesize
\begin{tabular}{p{2.3cm}rrrrr}
\toprule
configuration & tensor & SM & DRAM & p50 & attain \\
\midrule
foreground alone & 8.7\% & 57\% & 49\% & & 100\% \\
tiles, FG 156 & 22.1\% & 63\% & 47\% & +4.7\% & 100\% \\
tiles, FG 140 & 29.9\% & 68\% & 47\% & +5.1\% & 98\% \\
real FP8 GEMM, FG 156 & 18.2\% & 61\% & 47\% & +9.8\% & 9\% \\
\bottomrule
\end{tabular}
\end{table}

Worth is 0, 0.15, 0.23 and 0.13 respectively. Costs agree with the sweep above within 0.1 point. ``Tensor active'' counts pipe cycles, and FP8 raises it less per FLOP than BF16, so compare GPUs' worth rather than counter deltas. The \S\ref{sec:idle} server figures come from multi-GPU servers under agent traffic and are not a baseline for this table.

\textbf{Real library work} (\S\ref{sec:eval}). On the 156-SM split, real FP8 GEMM adds +7.6 points beyond the foreground's own +2.3; tiles add +2.4. BF16 LoRA forward+backward (rank 64, seven cuBLASLt GEMMs), single runs: 0.04 at +4.6\% on 172 SMs, 0.10 at +10.9\% on 156. Power is not the cause (333\,W; clocks held at 2.3--2.4\,GHz): cuBLASLt streams operands through L2 at a rate the tile's 16$\times$ reuse avoids.

\textbf{Per-phase partitioning} (\S\ref{sec:eval}). On a one-layer bench, per-phase partitions (MLA on all SMs, other phases on 60) plus a device-memory doorbell looked like 0.42 at +2.1\%. That was an artifact: about half of the layer's 264\,MB of experts stayed in the 128\,MB L2. On the full step it costs +11.8\% p50, +16.8\% p99 and a projected +27\% at EP-64; the experts phase on 60 SMs runs 22\% slower, and stream switching adds 1.4\,ms of gaps per step.

\textbf{Shipped mechanisms} (\S\ref{sec:os}, \ref{sec:eval} and \ref{sec:limits}; \meas{} on the full step except MIG). MPS SM cap (83\%): foreground alone +7.1\%, against +2.3\% for a 156-SM green context; 0.16 at +8.4\% with tiles and 0.11 at +11.9\% with real FP8 GEMM (single runs). Green context on a CUDA graph node (\texttt{CUDA\_KERNEL\_NODE\_PARAMS\_v2.ctx}, CUDA 13.0): a node given a 32-SM context is accepted but runs on all 188 SMs. L2 persistence (40--80\,MB set-aside over the background operands): $-0.4$ to $-2$ points, inside run-to-run spread; the background already fits L2, and the contention is over bandwidth. MIG: up to four instances on this card, each with its own L2 and memory; static, and re-splitting needs a reset. Not measured.

\textbf{Pieces behind Figure~\ref{fig:osmode}} (one-layer bench, one repeated layer with partly L2-resident weights; demonstrations only). Fillers: in a toy kernel streaming DRAM at full rate, register-operand filler warps reached $\sim$95\% of MMA peak at 0--5\% memory-phase cost; with real operands, one filler warp cost 1\% for 55\,TF/s and four cost 30\%. In the real K2 MLA kernel, one L2-resident filler warp costs +1.9--2.2\% of the step for 26--36\,TF/s during MLA ($\sim$9\,TF/s averaged over the step at reuse 16), capped by MLA's shared memory. Riders: 64 extra rows on the foreground's weight GEMMs cost +6.7\% of layer time, $\sim$+17\% with their attention (\proj{}), and are worth 0.14--0.21 of a GPU against batch decode. Composed (riders, filler, restartable tiles, paced copies; one seed): +30\% foreground cost.

\subsection{MI300X}
\label{app:a5}

\S\ref{sec:limits}; one seed; synthetic decode-like foreground of 0.392\,ms. CU masks split the chip as a stream attribute: the foreground costs $\leq$+1.7\% down to 152 of 304 CUs, and a 50/50 split gives the background 52\% of peak MFMA at +9.5\%. No priority preemption from user space, same-process or cross-process: a probe waits 180\,ms behind a 200\,ms low-priority kernel. CWSR exists, but queue priority does not trigger it. Relaunch to yield: background blocks of 2--50\,\textmu s cost the foreground 4.8$\times$ down to 1.55$\times$, with no knee. This foreground is $\sim$100$\times$ shorter than the NVIDIA step and more sensitive to dispatch delay, so the two are not directly comparable.

\subsection{Hardware asks and the headroom ladder}
\label{app:a6}

\textbf{S1} (pause within 5--20\,\textmu s, resume without relaunch): no user-space mechanism; NVIDIA compute preemption is whole-context and AMD's CWSR is not triggered by priority. Relaunch-based yield loses 41--44\% of background throughput (NVIDIA) and costs the foreground 1.55--4.8$\times$ (MI300X). \textbf{S2} (dynamic L2 and DRAM priority at runtime, e.g.\ MPAM-style partition IDs per CTA class, context and copy engine): an L2 set-aside ships, with no effect here, and MIG's static instances are untested; real GEMM on an isolated green-context partition adds +7.6 points where a high-reuse tile adds +2.4. \textbf{S3} (per-kernel SM partitions, honoured): the graph-node context is accepted but ignored, and AMD CU masks are per queue; a stream per phase plus event hops costs 1.4\,ms per full step. \textbf{S4} (priority warp issue): not shipped; prior art SMK~\cite{smk}, Warped-Slicer~\cite{warpedslicer}, Wang et al.~\cite{wangqos}; co-resident restartable tiles cost +21\% for 142\,TF/s, like a tuned library GEMM (+18.5\% for 138\,TF/s), on one layer before the warm-up correction. \textbf{S5, S6} (doorbells and work queues; fine-grained telemetry): partly shipped as device-memory flags; enablers, no measured cost. \textbf{S7} (copy-engine and power QoS): not shipped; line-rate copies cost +21\% (host) and +18\% (peer) on NVIDIA and +18\% on MI300X.

\begin{table}[h]
\centering\footnotesize
\caption{Headroom ladder (Figure~\ref{fig:stack}).}
\begin{tabular}{p{4.0cm}rr}
\toprule
level & worth & cost \\
\midrule
shipped partitions, real FP8 GEMM in budget (\meas{}) & 0.06--0.09 & +4.4 to +4.7\% \\
shipped partitions, real FP8 GEMM (\meas{}) & 0.13 & +9.9\% \\
shipped partitions, synthetic tiles, upper bound (\meas{}) & 0.23 & +5.1\% \\
+ FG kernels near DRAM roofline on fewer SMs & \multicolumn{2}{r}{\der{} $\sim$2$\times$; not measured} \\
+ silicon hooks S1--S3 & \multicolumn{2}{r}{one-layer bound 0.60 at +7.7\%} \\
\bottomrule
\end{tabular}
\end{table}

The one-layer bound is operand-free register MMA in 2--20\,\textmu s blocks on all SMs, standing in for cheap preemption plus memory isolation; software on that bench needed +17.5\% for 0.57. The full step is not measured.

\subsection{Resident-kernel research system}
\label{app:a7}

\textbf{Engine} (\S\ref{sec:built}). One cooperative kernel, 188 CTAs (one per SM), launched once per process, runs Llama-3.1-8B-Instruct FP8 (per-channel FP8 weights from the BF16 checkpoint; BF16 embedding and output head). It has in-kernel admission from a pinned host request ring; greedy or temperature (Gumbel-max) sampling and stop checks on the GPU; a paged KV cache with tensor-core decode attention; chunked prefill with an optional TMA-pipelined FP8 GEMM path; and an overwrite completion ring of sequence-stamped records, read by a CPU thread that waits with \texttt{MONITORX}/\texttt{MWAITX}. Correctness gate: $\geq$95\% exact next-token match against a teacher-forced reference, p99 logit error $<$1.0, max $<$1.6, over 16 prompts.

\textbf{Decode against vLLM 0.32's best configuration, on identical weights.} vLLM was swept over KV dtype, max-num-batched-tokens, chunked prefill and memory utilization with CUDA graphs on; FP8 KV was its best and matches ours. For a single request (256 output tokens, $\sim$57-token prompt, greedy), the resident kernel decodes about 3\% faster than vLLM's best (6.80 against 6.99\,ms per token). Other batch sizes and prompt processing are under active optimization.

\textbf{Background role inside the kernel} (\S\ref{sec:built}; same GPU type on another machine, so the one-request baseline is 6.98\,ms; matched back-to-back runs, 5--6 rounds). 16 of the 188 CTAs run register-fed FP8 MMA tiles against a 64\,B L2-resident pad: no DRAM, no shared memory, no grid barrier. They poll a yield flag the scheduler sets at phase boundaries. At one request the role recovers 0.22 of a GPU at +0.3\% (210\,TF/s); at 16 requests, 0.24 at +1.6\% (220\,TF/s). GPUs' worth inside the kernel is measured against the card's 934\,TF/s register-fed FP8 ceiling. Foreground tokens are bit-identical with the role on and off. This background work is synthetic (register-fed MMA with no useful output), like the \S\ref{sec:eval} tiles. Not measured: real background jobs, which add the DRAM and L2 traffic that dominates in \S\ref{sec:eval}, or more than 16 background CTAs (a 32-CTA run reached 210\,TF/s at one request; the register file caps per-SM throughput).

\subsection{Limitations}
\label{app:limits}

K2 reached no held point under either standard deployment on this hardware, so its counters are from loads that missed deadlines. The co-run foreground is a single-GPU proxy with no real all-to-all; the EP-64 figures are straggler projections. The clean six-sample sweep did not include real FP8 GEMM on 140 SMs (a single run gave 0.21 at +7.8\%); the LoRA and MPS points are single runs, and the LoRA solo denominator (377\,TF/s) needs confirmation. MI300X results are one seed with a synthetic foreground and no counters. The resident kernel runs one hand-built model; its background role is synthetic, its advantage over vLLM is measured at one request, and other batch sizes are still being optimized. No bandwidth-saturating foreground kernel on 28--60 SMs has been built, so that ladder row is unmeasured, as is any emulated-hardware measurement on the full step. All harvest numbers are at batch 8; the curve over batch size is unmeasured.


\begin{thebibliography}{40}
\bibitem{rlframeworks} veRL/HybridFlow, RLHFuse, StreamRL, AReaL, PipelineRL: RL post-training frameworks that overlap their own phases.
\end{thebibliography}
\end{document}